\documentclass[aps,superscriptaddress,twocolumn,showpacs]{revtex4}
\usepackage{epsfig}
\usepackage{amsbsy}
\usepackage{amsmath}
\usepackage{graphicx}
\usepackage{latexsym}
\usepackage{bm}

\newcommand\strikethru[1]{\leavevmode%
  {\setbox0\hbox{#1}\raise0.25\ht0%
   \hbox{\vrule height 1pt width\wd0\hskip-\wd0}\box0}}
\iffalse
  \newcommand\edit[2]{\def\first{#1}\def\second{#2}\def\empty{}%
                   {\ifhmode\ignorespaces{ }\fi
                    \bf\ifx\first\empty add: \else
                    {\bf\ifx\second\empty omit: \else change: \fi}
                             {\it#1}%
                     \ifx\second\empty\else{ \bf to: }\fi
                     \fi}{\bf#2}}
  \newcommand\comment[1]{{\bf #1}}
\else
  \newcommand\edit[2]{{#2}}
  \newcommand\comment[1]{}
\fi

\begin{document}

\title{Entanglement vs the quantum-to-classical transition}
\author{Shohini Ghose}
\affiliation{Institute for Quantum Information Science, University of Calgary, Alberta T2N 1N4, Canada}
\author{Paul M. Alsing}
\affiliation{Department of Physics and Astronomy, University of New Mexico, Albuquerque, NM 87131, USA}

\author{Barry C. Sanders}
\affiliation{Institute for Quantum Information Science, University of Calgary, Alberta T2N 1N4, Canada}

\author{Ivan H. Deutsch}
\affiliation{Department of Physics and Astronomy, University of New Mexico, Albuquerque, NM 87131, USA}

\date{\today}
\begin{abstract}

We analyze the quantum-to-classical transition (QCT) for
coupled bipartite quantum systems for which the position of one of the two
subsystems is continuously monitored. We obtain the
surprising result that the QCT can emerge concomitantly with 
the presence of highly entangled states in the bipartite system.
Furthermore the changing degree of entanglement is associated with the
back-action of the measurement on the system and is itself an indicator of the QCT.
Our analysis elucidates the role of entanglement in von Neumann's paradigm
of quantum measurements comprised of a system and a monitored
measurement apparatus.

\end{abstract}
\pacs{03.65.Ta, 03.65.Ud, 03.65.Yz, 05.45.Mt}
\maketitle


The quantum-to-classical transition (QCT) seeks to reconcile the quantum vs classical descriptions of the same phenomena.  Quantum and classical theories are distinguished both in terms of their state spaces and their dynamics.  Quantum states can predict measurement results that cannot be reconciled with predictions by classical states, such as violations of Bell's inequalities~\cite{Bell64}. Dynamically, although quantum and classical evolution agree on sufficiently short time scales (as a result of Ehrenfest's theorem~\cite{Sak}), the mean values of obervables diverge after some characteristic time, a break especially pronounced for chaotic systems~\cite{BB1979}. The importance of addressing the QCT in the context of both states and dynamics has been highlighted recently in the debate over the role of entanglement as a requirement for quantum computation~\cite{QC}. Here we analyze the nature of entanglement in states where the dynamics are classical.   By studying  a system coupled to a measuring apparatus as in the von Neumann seminal model~\cite{von32}, we define the QCT dynamically to be the condition that the measurement record evolves according to the classical equations of motion~\cite{BHJ2000}.  Surprisingly, this QCT is concomitant with a large amount of entanglement in the bipartite system. 

Following von Neumann's  quantum measurement model,  we consider a system $\mathcal{S}$ coupled to a continuously monitored measurement apparatus $\mathcal{A}$ treated as a single position degree of freedom (simply generalizable to multiple degrees of freedom).  We show (i)~the QCT, based on the dynamical measurement record for $\mathcal{A}$, may concomitantly be accompanied by a high degree of entanglement in $\mathcal{S}$+$\mathcal{A}$ (the joint system and apparatus), (ii)~demonstrate that the QCT can be effectively characterized by the width of the distributions for the measured quantities and the changing degree of entanglement, and (iii)~provide an example of a coupled harmonic oscillator and spin system as an example of the coexistence of strong entanglement and the QCT. Our analysis elucidates the role of entanglement in von Neumann's measurement paradigm, even when the acquired data from measurement of $\mathcal{A}$ agrees with classical data.   It does not restrict the nature of~$\mathcal{S}$, and as we assume continuous measurement, it is relevant to realistic models in a variety of contemporary experiments~\cite{Exp}.

A general quantum dynamical process is described by a completely positive map $M:\rho\mapsto\rho'$ for $\rho$ the initial density operator and $\rho'$ the final density operator. We regard~$M$ as exhibiting a QCT transition if the measurement record obtained from monitoring~$\mathcal{A}$ is increasingly indistinguishable from a record predicted by classical dynamics as $\hbar \rightarrow 0$, or more precisely, as the dimension of the required Hilbert space approaches infinity. In this limit, we explore the non-classical nature of the state $\rho$ by examining the entanglement between $\mathcal{S}$ and $\mathcal{A}$. When~$\mathcal{A}$ is continuously monitored in a weak measurement, the measurement record is describable by a quantum trajectory~\cite{Car93}.  This description is intrinsically stochastic with irreducible noise arising from the information gain/back-action trade off.  
For ideal measurements (which acquire all information that leaves $\mathcal{S}$+$\mathcal{A}$), the state~$\rho$ of $\mathcal{S}$+$\mathcal{A}$ is always pure, with the specific pure state differing according to the corresponding measurement record. 
The degree of bipartite entanglement can be obtained equivalently as the entropy $S(\rho_\mathcal{S})$ or $S(\rho_\mathcal{A})$ for $\rho_\mathcal{S}=\text{Tr}_\mathcal{A}\rho$
and $\rho_\mathcal{A}=\text{Tr}_\mathcal{S}\rho$. We work with the linear entropy
$S=1-\text{Tr}(\rho_\mathcal{S}^2)=1-\text{Tr}(\rho_\mathcal{A}^2)$,
which is convenient to calculate and can be employed as an estimator for other required entanglement measures~\cite{Ber03}. 

Consider how $\rho$ behaves under mappings that obey the QCT.  The stochastic evolution of the pure density matrix for $\mathcal{S}$+$\mathcal{A}$ is given by
\begin{align}
\label{SMEc}
	\text{d}\rho=&-\frac{\text{i}}{\hbar}[H,\rho]\text{d}t+k(2q\rho q -q^2\rho - \rho q^2)\text{d}t
			\nonumber\\&+\sqrt{2k}(q\rho+\rho q -2\langle q \rangle \rho)\text{d}W
\end{align}
with measurement record $\text{d}X=\langle q \rangle \text{d}t + (8k)^{-1/2}\text{d}W$ 
on $\mathcal{A}$~\cite{Measurement}, with~$k$ the strength, (resolution) of the position measurement
and d$W$ the Wiener noise.
Entanglement of the $\mathcal{S}$+$\mathcal{A}$ state is quantified by the linear entropy of the reduced density operator
for either~$\mathcal{S}$ or~$\mathcal{A}$, and we choose to follow the evolution
of the reduced density operator for~$\mathcal{A}$ as it is known to have one degree of freedom,
namely position. The stochastic evolution of the reduced density operator for~$\mathcal{A}$ is
\begin{align}
\label{SMEc2}
	\text{d}\rho_\mathcal{A}=&-\frac{\text{i}}{\hbar}\text{Tr}_\text{S}([H,\rho])\text{d}t
		+k(2q \rho_\mathcal{A} q -q^2  \rho_\mathcal{A} -  \rho_\mathcal{A} q^2)\text{d}t\nonumber\\
	&+\sqrt{2k}(q \rho_\mathcal{A}+ \rho_\mathcal{A}q -2\langle q \rangle \rho_\mathcal{A})\text{d}W.
\end{align} 
The evolution of the marginal linear entropy for~$A$ obeys
$\text{d}S=-2\text{Tr}(\rho_\mathcal{A}\text{d}\rho_\mathcal{A})-\text{Tr}(\text{d}\rho_\mathcal{A}^2)$.
From Eq.~(\ref{SMEc2}) (and only retaining terms to $\text{O}(\text{d}t)$),
we obtain an expression for the evolution of the entanglement for a given measurement strength $k$,
\begin{align}
\label{dS}
\text{d}S_k = \text{d}S_0&-8k(\langle q\rho_\mathcal{A} q \rangle 
 	-2\langle q\rho_\mathcal{A}\rangle\langle q\rangle
	+\langle q \rangle^2  \langle\rho_\mathcal{A}\rangle  )\text{d}t \nonumber\\
	&- 4k (\langle q  \rho_{\mathcal{A}}\rangle  -\langle q \rangle \langle \rho_{\mathcal{A}}\rangle)\text{d}W
\end{align}
with $\text{d}S_0$ the term corresponding to measurement-free ($k=0$) evolution.

To consider the QCT, we study the evolution of the moments of position and momentum of the measured subsystem~$\mathcal{A}$.
These equations of motion are
 \begin{align}
\text{d}\langle q \rangle &= (\langle p \rangle /m)\text{d}t+\sqrt{2k}C_{qq} \text{d}W,\nonumber\\
\text{d}\langle p \rangle &= -\langle \partial_qV\rangle \text{d}t+\sqrt{2k}C_{qp} \text{d}W
\end{align}
with centroid coordinates $\langle q \rangle$ and $\langle p \rangle$
and covariances $C_{ab}=({\langle a b\rangle
	+\langle ba \rangle})/{2}-\langle a \rangle \langle b \rangle$.
The force $-\partial_q V$ acting on~$\mathcal{A}$ is derived from the potential~$V$.
Provided that the noise terms (due to back-action) are sufficiently small
and the measurement is sufficiently strong to localize the state, these equations of motion will closely follow
the dynamics of the corresponding classical system: the strong localization and weak noise
conditions require that the variances and covariances remain small compared to the phase space explored by
the classical equations of motion~\cite{BHJ2000}.  The position measurements act to damp correlations beyond second order.

In this limit  (covariances are small), the measurement terms in Eq.(\ref{dS}), which can be written in terms of the  covariances,
become negligible and the degree of change in entanglement, 
\begin{equation}
\label{eq:Sk}
\Delta S_k= \sup_t \left(1-\frac{S_k(t)}{S_0(t)}\right),
\end{equation}
caused by observation of~$\mathcal{A}$, approaches zero. 
Since $\Delta S_k$ is negligible only if the covariances, which are proportional to the back-action noise terms in Eq. (5) become sufficiently small, it quantifies the degree of back-action resulting from the  measurement of~$\mathcal{A}$ with a given measurement strength~$k$. Hence $\Delta S_k \rightarrow 0$ coupled with the localization
condition discussed above, are sufficient conditions to ensure the QCT for this bipartite system.

In the QCT, $\Delta S_k$ must remain small, but no such restriction applies to $S_k(t)$ itself.
For example a quantum chaotic system may enhance the bipartite entanglement
of $\mathcal{S}$+$\mathcal{A}$ more rapidly than the measurement process can diminish the entanglement,
and therefore large bipartite entanglement may be compatible with the QCT.
The compatibility of large entanglement and the QCT for the dynamics is illustrated by
the following example.

The system + apparatus~$\mathcal{S}$+$\mathcal{A}$ consists of a particle of mass $m$ in a harmonic trap of
angular frequency $\omega$.
The particle is  coupled via an internal magnetic moment to a gradient magnetic field along its 
axis of motion~$z$ and a constant transverse field along~$x$.
The corresponding Hamiltonian governing the dynamics is
\begin{equation}
  H = \frac{p^2}{2m}+ \frac{1}{2}m\omega z^2 + b  z J_z+cJ_x,
    \label{H}
\end{equation}
which applies to various phenomena including the simplest Jahn-Teller model~\cite{Eng72},
the Jaynes-Cummings model without the rotating wave approximation~\cite{Jay63}, 
and the motion of ultra-cold atoms in a magneto-optical trap~\cite{DAGGHJ2000}.
The classical Hamiltonian has the same form as Eq.~(\ref{H}) with the $z$ and $p$ operators replaced by classical variables and the spin replaced by a classical magnetic moment. The transverse magnetic field along $x$ causes the Hamiltonian to become non-integrable and leads to chaotic dynamics~\cite{Chaos}.  Previous studies showed that a continuous position measurement resulted in quantum trajectories that exhibit classical chaos when the actions associated with the spin and harmonic motion are both large relative to $\hbar$~\cite{GADBHJ2003, GABHD2003}. Here we analyze the behavior of entanglement in this limit. 

We introduce initial states that are products of gaussian and spin coherent states $|\psi(0)\rangle=|\alpha\rangle|\theta,\phi\rangle$. We start with a spin of $J=200\hbar$ which puts us in the semiclassical regime. We let $c=$ and $b=m\omega^2\Delta z/J$ with $\Delta z = 45z_g$ where $z_g=\sqrt{\hbar/2m\omega}$. This results in a characteristic external action $I_0=m\omega\Delta z^2 = 1000\hbar$.  Classical trajectories are recovered for a measurement strength of $k=\omega/8z_g^2$ given these parameter choices~\cite{GABHD2003}. 

\begin{figure}[htbp]
\centerline{{\epsfig{file=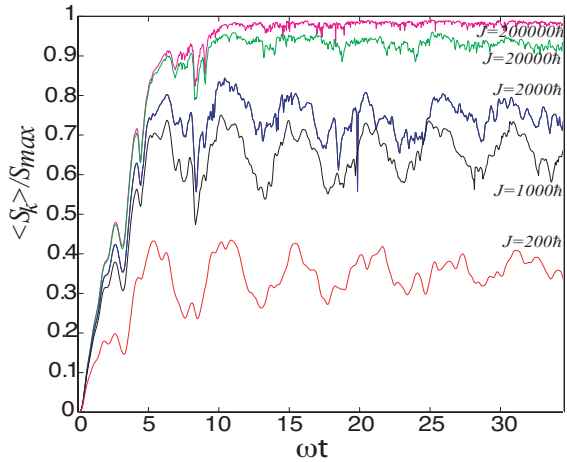,width=75mm}}}
\caption[Evolution of the entanglement]{(a) The normalized average linear
entropy increases as J is increased, keeping $I_0/J$ constant. }\label{LEChaos}
\end{figure}

Figure~\ref{LEChaos}  shows the evolution of the average normalized linear entropy 
$\langle S_k \rangle/S_\text{max}$ of 100 trajectories with $S_{\text{max}}=1-1/(2J+1)$ and energy $E=0.58E_0$ ($E_0=m\omega^2\Delta z^2$) for increasing values of $J$ (starting with $J=200\hbar$), keeping the measurement strength $k$ constant. We scale $b$ up appropriately relative to $z_g$ in order to keep the ratio $I_0/J$ constant in the classical limit. As both $J$ and $I_0$ are increased and the measured quantum trajectories approach the classically predicted trajectories~\cite{GABHD2003}, the average entanglement increases.  This indicates that in a regime where classical trajectories emerge from the measured quantum system, the underlying states are highly entangled.

This behavior can be understood by examining the  measured state more closely. If we write the state $\vert\phi\rangle$ in terms
of its spin components $\vert m \rangle$ in some basis as
\begin{equation}
\vert \phi \rangle = \sum_{m=-J}^{J} 
               \alpha_m \vert \phi_m  \rangle \vert m \rangle,
\end{equation}
then we can relate the entanglement between spin and motion  to the overlap between the spinor components in the different spin states:
\begin{equation}
S_k = 1 -  \sum_{m,n} \left|\alpha_m^* \alpha_n \langle \phi_m  \vert \phi_n  \rangle\right|^2.
\end{equation}
If there is zero overlap between different spinor components, then the only contributions to the sum are for $m=n$. In this case, when in addition the $\alpha_m$ are all equal, the state is maximally entangled.

In order to understand the behavior in the classical limit, we first consider  the spin-$\tfrac{1}{2}$
case studied in~\cite{GADBHJ2003} with $c=0$.  The spinor components in the diabatic ($|m_z\rangle$) basis of an initial spatially localized state move along two different harmonic wells centered at $\pm b$  so that their overlap reduces almost to zero (Fig.~2(a)). At this point, the entanglement increases to its maximum value, but falls back to zero when the measurement eventually projects the state into one of the two spin states. Thus, in this example maximum entanglement (zero overlap between wavepackets) corresponds to a measurement that perfectly distinguishes the spinor components, resulting in a projective measurement with  maximum measurement back-action. The entanglement acts as a measure of the noise on the spin due to the position measurement.

In the large action chaotic limit, the increase in $S_k$ with the actions  can be understood in a similar manner. The overall extent of the initial state in position and momentum  spreads as the $2J+1$ spinor components move along different diabatic potentials
and are coupled by the transverse magnetic field. As the spinor components spatially separate, their overlap decreases, leading to an increase in entanglement. At the same time, the measurement acts to localize the state, thereby damping out the tails of the spatial distribution and preventing further spatial separation between the spinor components. 
Whereas in the spin-$\tfrac{1}{2}$ case, the measurement was strong enough to eventually resolve the two spinor components, for larger spin~$J$, the same measurement strength cannot distinguish between all $2J+1$ components (Fig.~2(b)). Thus the measurement does not  project  the state into one of the $2J+1$ spin states and hence the entanglement does not decrease back to zero. Instead, a quasi-steady state is reached where some non-overlapping spinor components remain that  lead to a non-zero steady-state entanglement, but are indistinguishable by the measurement. For a constant measurement strength  $k$, as the actions~$I_0$ and~$J$ increase, more non-overlapping spinor components fit within the width of the measurement resolution, and hence this steady state entanglement increases as seen in Fig.~\ref{LEChaos}. 

\begin{figure}[htbp]
\centerline{{\epsfig{file=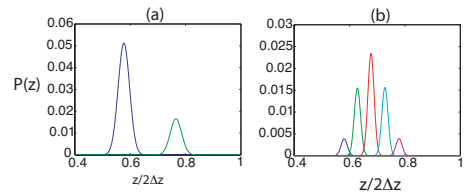,width=60mm}}}
\caption{(a) In the spin-$\tfrac{1}{2}$ system, the measurement is
strong enough to resolve the two components of the wave function and thus remove the entanglement. (b) For larger actions even though the wave packet components are distinguishable so that the state is highly entangled, the same measurement is too weak to resolve all the different wavepackets and remove all the entanglement (b).}
\end{figure}
 
Whereas the weak measurement does not remove all the entanglement between spin and motion, it is {\em sufficient} to localize the state and damp
the higher order cumulants that lead to ``nonclassical'' dynamics~\cite{GADBHJ2003,GABHD2003}.  The connection between entanglement and the cumulants becomes clear in the large action (classical) limit, where the measurement causes the reduced state of the motional subsystem to remains approximately Gaussian.  In that case the linear entropy can be written solely in terms of the variances and covariance as
\begin{equation}
S_{\text{Gauss}}=1-{{{\hbar  \mathord{\left/ {\vphantom {\hbar  2}} \right.
\kern-\nulldelimiterspace} 2}} \over {\sqrt {C_{zz}C_{pp}-C_{zp}^2}}}.
\label{LEGaussian}
\end{equation}
The quantity $\sqrt{C_{zz}C_{pp}-C_{zp}^2}$ represents the effective area of the ``uncertainty bubble" of the Gaussian distribution and its ratio to $\hbar/2$ measures the number of minimum uncertainty wave packets which fit within this area and thus the dimension of the Hilbert space required to describe the marginal state.  This ratio therefore determines the effective rank of the reduced density operator, or Schmidt number of the entangled state.  From this equation it is clear that even if the variances and covariances remain small relative to the {\em total phase space} of the dynamics, they may still be large compared to~$\hbar$, and the entanglement can be close to maximal ($S \approx 1$). In this way, one can simultaneously satisfy the QCT conditions  (covariance matrix remains bounded), thereby
acquiring a measurement record predicted by classical Hamiltonian equations and obtaining
an evolution that results in  a highly entangled quantum state.  This arises because the various $J_z$ components are {\em {in principle}} distinguishable by a position measurement alone, provided that the measurement can resolve the separation between the packets.  However, classical mechanics emerges precisely because {\em{in practice}} the measurement is weak and hence cannot induce strong quantum back-action.
\begin{figure}[htbp]
\centerline{{\epsfig{file=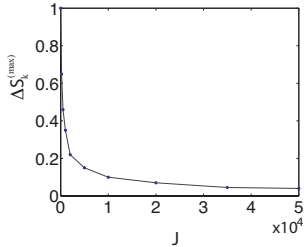,width=40mm}}}
\caption{$\Delta S_k$ decreases as $1/\sqrt{J}$ for a fixed measurement strength $k$.}
\end{figure}

The measurement back-action can be quantified by $\Delta S_k$. Consider first the extreme quantum limit, $J=\tfrac{1}{2}$.  In that case, any measurement strong enough to localize the wavepacket in position will necessarily resolve the two spinor components and cause {\em maximum back-action},  projecting the maximally entangled state on to one of the two spin states. This is  accompanied by a maximal change in degree of entanglement, $\Delta S_k$  from its maximum value (corresponding to a ``Schr\"odinger cat" state) to zero (a product state). In contrast, in the large action limit, the  measurement is only weakly projective (small back-action) on the spin system and correspondingly, does not change the entanglement as much. If we replace $S_0$ by $S_{\text{max}}=1-\tfrac{1}{2J+1}$, we obtain an upper bound on $\Delta S_k$. Figure 4 shows the steady-state behavior of this upper bound as a function of the size of the spin system. As the system is made more classical by increasing $J$ and $I_0$, keeping $k$ fixed, the maximum value of $\Delta S_k$ decreases rapidly, indicating that the back-action due to the measurement 
decreases, as is expected in the classical limit. On the other hand, keeping the actions fixed, as $k$ increases, $\Delta S_k$ increases, reflecting the larger back-action on the spin system caused by a stronger measurement of the position. $\Delta S_k$ is thus a good quantitative measure of the small back-action condition required for the quantum to classical transition and provides an alternative to the covariance matrix conditions obtained in~\cite{GABHD2003}. For a given measurement strength $k$ which sufficiently satisfies the localization conditions, classical dynamics will be recovered if $\Delta S_k$ also remains small.

In conclusion, we have examined the QCT in continuously measured coupled systems and revealed the surprising result that even when classical dynamics can be recovered in the measured trajectories
for the apparatus~$\mathcal{A}$, the underlying states may be extremely non-classical exemplified
by a large amount of bipartite entanglement.  The key point is that the measured system will approximately follow classical trajectories when the cumulants are a small fraction of the total classical phase space measured by some characteristic action.  In contrast, entanglement depends on the growth of these cumulants with respect to the {\em absolute} scale of action, $\hbar$.  Thus, as the action increases and one moves into the classical domain where a macroscopic phase space is explored, the relative size of the cumulants decreases while the entanglement increases. In such a regime of large entanglement, the QCT can be recovered if the condition of small change in entanglement $\Delta S_k$ is fulfilled along with the strong localization condition. Thus entanglement can be used to quantitatively identify the QCT in coupled systems. 
The QCT for coupled systems is especially interesting in the standard paradigm of quantum
measurements, introduced by von Neumann, comprising a system~$\mathcal{S}$ and a 
monitored, or observed, measurement apparatus~$\mathcal{A}$: our results demonstrate that 
continuous monitoring of~$\mathcal{A}$ can yield an ostensible QCT (the measurement record
is consistent with classical theory) and concomitant strong bipartite entanglement of the 
$\mathcal{S}$+$\mathcal{A}$ state.
In addition to elucidating the QCT, our results are also of relevance to 
the proliferation of experiments in which individual quantum subsystems can now be continuously monitored. 

We appreciate valuable discussions with S. Habib and T. Bhattacharya. This project has been supported by Alberta's Informatics Circle of Research Excellence (iCORE), the Alberta Ingenuity Fund and the National Science Foundation under  Grant No. 0355040.

\end{document}